\documentclass[12pt]{article}
\usepackage[utf8]{inputenc}
\usepackage{setspace}
\usepackage[english]{babel} 
\usepackage{amsmath,amssymb,latexsym}
\usepackage{slashed}
\usepackage{graphicx}
\usepackage{epstopdf}
\usepackage{mathtools}
\usepackage[sorting=none, backend=bibtex]{biblatex}
\bibliography{./refs.bib}

\usepackage{svg}

\usepackage{xcolor}
\usepackage{hyperref}

\usepackage[margin=1in]{geometry}

\usepackage{authblk}

\newcommand{\myverb}[1]{{\small\texttt{#1}}}

\def \sha {\myverb{SHA256}}
\def \Scramble {\myverb{Scramble}}
\def \Create {\myverb{Create}}
\def \Verify {\myverb{Verify}}
\def \beq {\begin{equation}}
\def \eeq {\end{equation}}

\title{Black holes and cryptocurrencies}
\author[1]{Alexey Milekhin\footnote{Email: milekhin@ucsb.edu}}
\affil[1]{Department of Physics, University of California, Santa Barbara, CA 93106, USA}
\date{\today}
\begin{document}

\maketitle

\begin{abstract}
It has been proposed in the literature that the volume of Einstein--Rosen bridge is equal to complexity of state preparation ("Complexity=Volume" conjecture). Taking this statement outside the horizon, one might be tempted to propose "Complexity=Time" correspondence. 
    In this Essay\footnote{Essay written for the Gravity Research Foundation 2022 Awards for Essays on Gravitation.} we argue that in  a blockchain protocol, which is the foundation of all modern cryptocurrencies, time is emergent and it is \textit{defined} according to a version of "Complexity=Time".
\end{abstract}

\section{Introduction}
\begin{figure}[!h]
\centering
\begin{minipage}{0.45\textwidth}
\includesvg[scale=0.8]{./ER_bridge2.svg}
\end{minipage}
\begin{minipage}{0.45\textwidth}
\includegraphics[scale=1.]{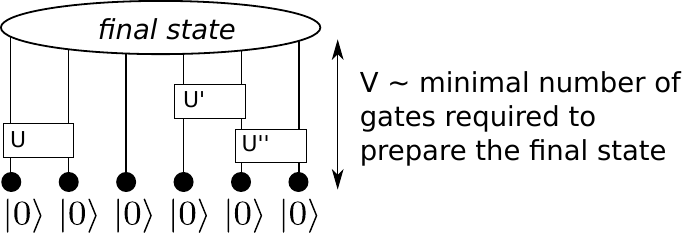}
\end{minipage}
\caption{\textit{Left:} ER bridge and its growth. \textit{Right:} quantum circuit.}
\label{fig:bh}
\end{figure}

Black hole interior is a mysterious place. Celebrated Einstein--Rosen(ER) bridge 
solution has two universes connected via a two-sided black hole interior - Figure \ref{fig:bh} (Left). The volume of this interior, evaluated on certain Cauchy slices, grows linearly 
with time. It is expected that this growth can continue until the volume becomes of order $e^{S_0}$, where $S_0$ 
is the black hole entropy defined as the quarter of its horizon area. 

If we believe that $S_0$ characterizes the number of black hole degrees of freedom, 
the volume can become exponentially big in this number. However, all usual thermodynamic 
quantities scale linearly with the number of degrees of freedom, not exponentially. It prompted the authors of \cite{susskind2016computational, brown2016complexity, brown2018second} to argue 
that the volume of the ER bridge is equal to the (quantum) complexity of state preparation:
\beq
\textit{Complexity=Volume.}
\eeq
This complexity is defined as a minimum number of “basis” quantum 
gates required to prepare a state starting from some fixed simple one - Figure \ref{fig:bh} (Right).
This quantity can be as big as $e^{\text{number of qubits}}$, but cannot be bigger. 
Recently it was explicitly demonstrated \cite{luca} in a model of 2d gravity that ER-bridge length 
indeed saturates at $e^{S_0}$.
We refer to \cite{Couch_2017,python,python2,anything} regarding the recent challenges to this conjecture.

``Complexity=Volume'' operates with the volume inside a black hole interior. 
Colloquially, when go outside a black hole interior we trade a space coordinate to a time coordinate.
Hence more generally, without any black holes, we can expect that 
\beq
\textit{Complexity=Time.}
\eeq
It would be very interesting to find explicit physical examples of such correspondence. 
The closest analogy known to the author is the construction by Page and Wootters \cite{PageWootters},
where non-trivial time evolution in energy eigenstates manifests in non-trivial conditional
expectation values of local operators \footnote{To put it differently, an expectation
value of an operator can serve as a clock.}. 
It has been proposed before that spacetime could arise from entanglement \cite{Swingle,Verlinde,Van,EREPR}. However,
standard quantum entanglement is a feature
\footnote{There are certain notions of "entanglement in time"\cite{TE1,TE2,TE3}. It has been proposed to 
use it in order to create a quantum analogue of blockchain \cite{rajan}. In this Essay we concentrate on
classical blockchain.}
 of a given state(i.e. Cauchy slice). We propose that 
time dynamics comes from complexity.

In absence of “conventional” physical examples,
in the rest of
this Essay we would like to argue that ``Complexity=Time'' is the underlying principle of a blockchain
protocol, which is the foundation of all modern cryptocurrencies. We will see a few familiar physical
phenomena in blockchains:
\begin{itemize}
\item Lorenzian time-like geodesics maximize the proper distance. Correspondingly,
the ``current time'' in blockchains is defined by finding the maximal complexity slice.

\item Non-traversability of ER-bride can be attributed to its length growth: the length grows
too fast for an observer to cross the horizon. In blockchain, a single observer(with limited
computational resources)
cannot ``change history''(i.e. time travel) exactly for the same reason: the blockchain grows too fast. 

\item As a bonus point, blockchain has a version of error correction: communication errors between
individual observers do not cause the system to break down as a whole.

\item It was proposed by Hayden and Preskill \cite{HP}, that once a 
black hole has radiated away half of its entropy, an observer who has 
access to the radiation can actually retrieve information from the interior. In blockchain
\footnote{I am grateful to Ying~Zhao for reminding me about Hayden--Preskill protocol.}
, an observer
who has $(50+\epsilon)\%$ of the total computational power can in fact change the entire history of
blockchain. We can interpret this as an ability to actually cross the horizon to get information about
the interior despite interior growing very fast. However, rewriting blockchain history will still
require a lot of computational time. This agrees with results of Harlow and Hayden \cite{HH}. 

\end{itemize}
Although a reader should keep in mind that the discussion in the next Section is purely classical.
Whenever we refer to "complexity" in blockchain it is (appropriately defined) classical complexity of
performing arithmetic operations.

\section{Essentials of blockchain}
In this Section we describe the essential properties of blockchain using physics language. The
presentation here includes only the most basic features. For example, we will omit public key
signatures.

In physics terms, blockchain solves the problem of emergent (decentralized) time:
\begin{itemize}
    \item Suppose we have $N$ observers who constantly produce and send each other classical messages
about events happened to them.
How can they agree on a global time? 
By "time" here we mean arranging the messages they sent/received in a linear order. 
\end{itemize}
Moreover, most of the observers should end up with the same sequence of messages. Synchronizing clocks and then 
time--stamping messages is not a good solution: clocks can easily be desynchronized (either randomly
or at will) and 
observers would not be able to resolve disputes. If Alice and Bob have 
messages in different orders they would not be able decide between themselves who is right.
Moreover, one observer may send different versions of the 
same message to different people. 
This way different observers may record different history and there is no consensus between 
them.
In computer science language one may call this “attack” or
“malignant intent”, but in physics language we can attribute this to a statistical
fluctuation. 

Let us illustrate this point. Suppose we have three observes: Alice, Bob and Carol. Carol wants to 
divide the inheritance between Alice and Bob. A problem might arise if she sends “Alice gets everything”
to Bob and “Bob gets everything” to Alice:
\beq
\includegraphics[scale=0.6]{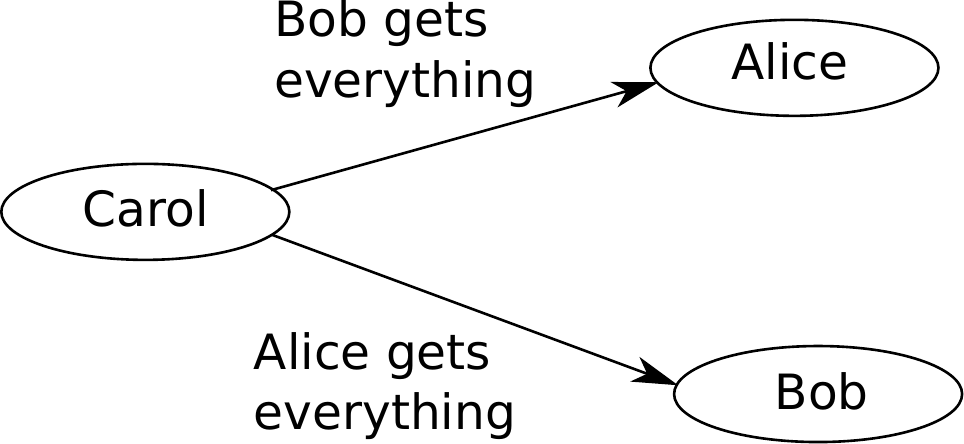}
\eeq
How do Alice and Bob decide which version is right?

Another problem may arise if Carol initially declares that Alice should inherit everything, but then
changes her mind and sends a message to everyone that Bob gets everything:
\beq
\includegraphics[scale=0.6]{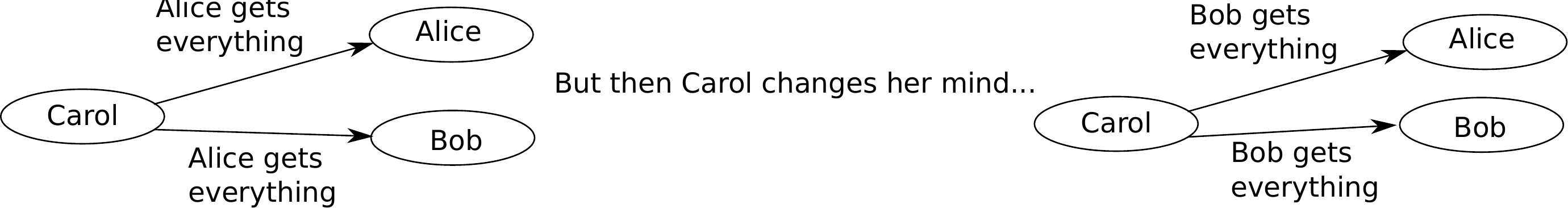}
\eeq
This is equivalent to travelling back in time to change the will. 

Blockchain \cite{nakamoto} solves both of these problems under a number of reasonable assumptions. We have to assume that
observers have comparable computational resources and that there is a bound on how long it takes to 
send a message between the observers (we refer to \cite{miller} for a discussion on this point). 
It can be arbitrary big, but it has to be bounded.

One may wonder what does it mean to have “comparable computational resources” and “how long it takes
to send a message” if we do not have a global time. Is it not the point of this paper to 
use the computational complexity to define “time”? The most important thing here is that we
have multiple observers. For a single observer who, by whatever reason, is capable of 
producing messages by doing mathematical operations there is no problem at the first place: the sequence
of their messages can define time for them. The problem arises when we have multiple observers.
If we bring them together they can compare the rates at which they can do computations. They
can do it in terms of “how many ”standard“ computations my neighbor can do while I 
do 1 ”standard” computation“.
They can measure distance in a similar way. As a bonus point, we will allow these two quantities to
have finitely-bounded fluctuations. From General Relativity we know that we cannot synchronize clocks
in this case to define a truly global time. This is why this whole construction will be probabilistic.

First of all, blockchain forces a one-dimension structure on the sequence of messages.
We want to arrange messages in a chain of blocks, hence the name blockchain:
\begin{equation}
\includesvg{./blockchain.svg}
\end{equation}
We need appropriate functions to create new blocks (local step) and we 
need to somehow compare different chains (global step). 
The first involves the so-called proof-of-work \cite{dwork}.
We introduce two functions: \Create, which creates a valid new block given the previous one and some message
we want to store:
\beq
\text{Block N} = \Create(\text{Block N-1}, \text{new message}),
\eeq
and \Verify, which verifies the validity of block $N$ given the previous block:
\beq
\Verify(\text{block N}| \text{block N-1}) = \text{True or False}.
\eeq
The important distinction between these two functions is the
computational complexity: creating(mining) a new block 
must be very computationally expensive, 
whereas verifying the validity of a block must be very easy
\footnote{
The way it is achieved in Bitcoin is the following. A single block has the following
structure:
\begin{equation}
\text{new block:} \quad
    \begin{tabular}{|c|}
         \Scramble(\text{the whole of previous block})   \\
         \hline
         \text{new message}  \\ 
         \hline
         \text{nonce}
    \end{tabular}
\end{equation}
Each message is, of course, the amount of money one user sent to another. In this sense
any cryptocurrency is just a public ledger.
Function \Scramble \ is a highly-scrambling function. It is easy to apply it, but very hard to invert.
It would be convenient to assume that it produces a number.
In computer science they are often referred to as one-way functions. For example, Bitcoin uses \sha \ as a one-way function. It can map anything to a 256-bit number. Examples of its outputs(using 16-base numbers) are: \newline
\sha("Hello World")=64ec88ca00b268e5ba1a35678a1b5316d212f4f366b2477232534a8aeca37f3c \newline
\sha("Hello, World")=4ae7c3b6ac0beff671efa8cf57386151c06e58ca53a78d83f36107316cec125f \newline
“Nonce” is just an extra “junk” number which must be added to satisfy the following validity 
criteria(we are still simplifying a few things): \newline
$$
\text{Number \Scramble(whole new block) starts with 8 zeros.}
$$
In general, there are no good ways to produce the correct nonce except bruteforce trial and error.
An attempt to change a message inside the block number $N-1$ will lead to a 
different \Scramble(block N-1) and hence one will have
to find a new nonce for the block number $N$.
}.

Now we know how to create chains of blocks. How do we compare different chains? Satoshi Nakamoto in their
paper on Bitcoin proposed the following criteria \cite{nakamoto}:

\beq
The\ longest\ (valid)\ chain\ is\ the\ correct\ one.
\eeq
Here the word ``longest'' can be substituted by ``the most complex'', as each block has approximately
the same complexity. So if an observer wants to append a block, they should find the longest (valid) chain and
append their block there. This way blockchain constantly grows.

One important exception is when we have two chains of equal length (a fork). Let us return to
the example of Carol sending different messages to Alice and Bob. Now that the
sequence of messages is one dimensional, she will have to create a fork where in one
branch Alice gets everything and Bob gets everything in the other:
\beq
\includegraphics[scale=0.9]{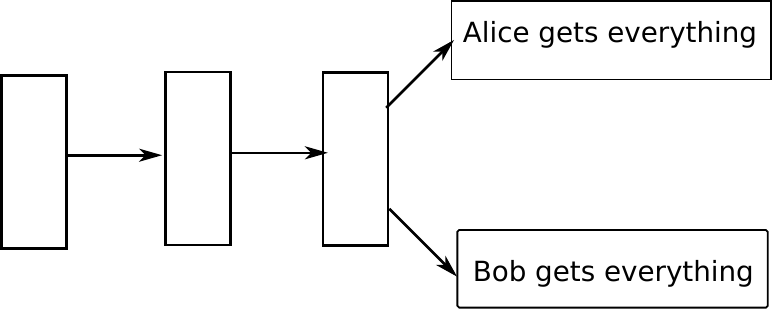}
\eeq
In this
case a third-party observer who sees these two chains can pick up any branch they want to continue
attaching messages. Other users will do that too and after a few more blocks
the tie will be broken and one
of the branches will dominate. This is illustrated by the following figure:
\beq
\includegraphics[scale=0.9]{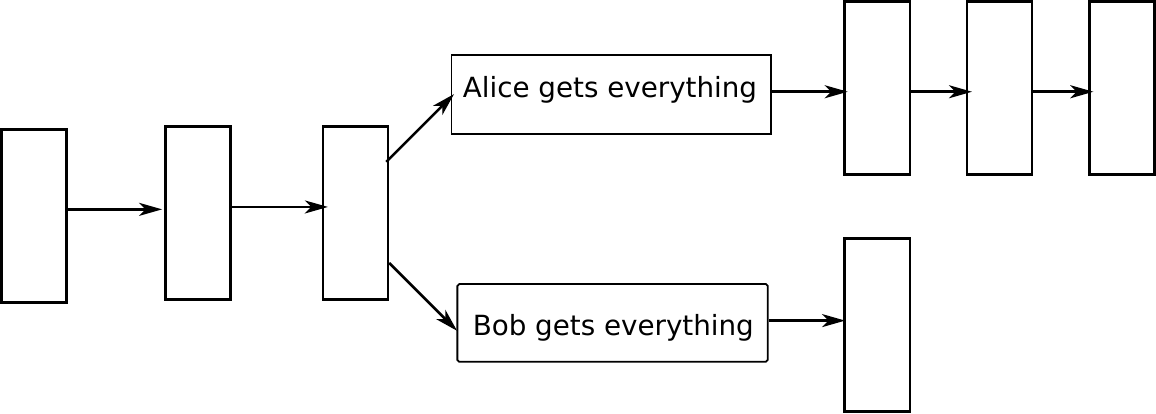}
\eeq
\textit{It is in this sense that blockchain has error correction: possible error on Carol's side
does not lead to loss of consensus among other observers.}

However, one must be aware that it introduces a probabilistic aspect in a history defined by 
blockchain: a newly created block(blue) might not stay in the (dominant branch of) blockchain, 
as it might not reach
other observers before they extend the blockchain differently(green),
rendering the original block to be an outcast:
\beq
\includegraphics[scale=0.7]{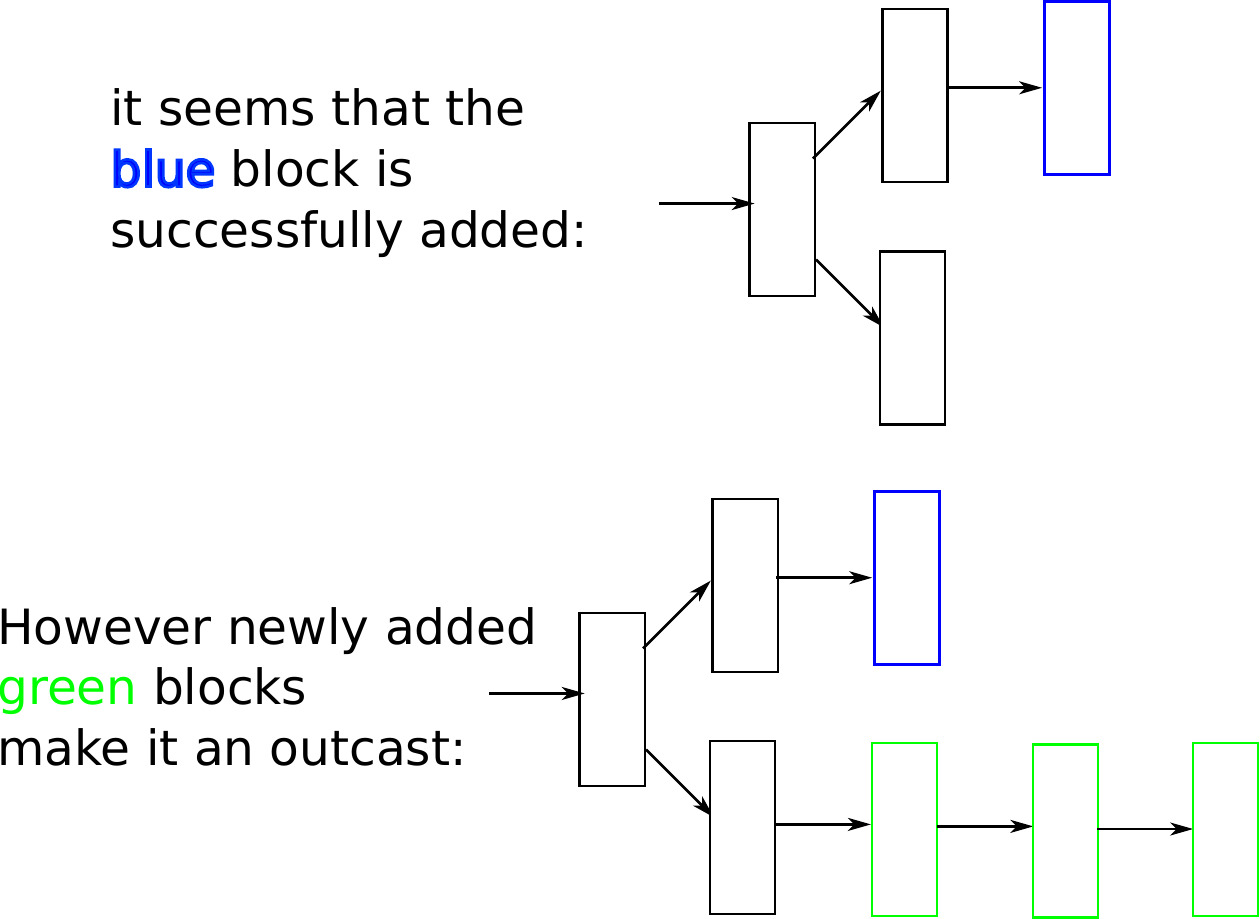}
\eeq

So the resulting structure resembles a tree rather than a
chain: there are many blocks(red) which do not belong to the maximal complexity slice:
\beq
\includesvg[scale=0.7]{./tree.svg}
\eeq
\textit{The maximal complexity slice determines which events actually happend and in what order they
happened.}
Given a block, the more blocks there are after it the more likely it will stay on the maximal slice\footnote{In Bitcoin, after adding a new block it is standard to wait for an extra 3-4 new blocks to make 
sure your block will stick.}.
The block number along the maximal complexity slice is what one might call ``emergent time''.
This is similar to time-like geodesics in Lorenzian signature, as they too are supposed to have maximal
length.


Let us see show blockchain prohibits time travel. Suppose Carol wants 
to go back in time to change her will. She will have to change the corresponding block
in the far past (in blockchain time), as at present Alice owns everything.
She cannot just modify a single block in the past, as it will render the subsequent blocks invalid:
remember that \Verify \ function depends on a block and its parent. “Time-traveller” will have to 
create a new branch and
redo the whole sequence of blocks after the modified one.
The newly created branch will be rejected by other observers, as it has low complexity. With time
the disparity will only grow, assuming that more than a half of observers work on the main branch. It is
very computationally expensive to produce new blocks and the main branch will grow much faster.
This is illustrated by the following figure (red blocks represent the alternative history):
\beq
\includegraphics[scale=0.7]{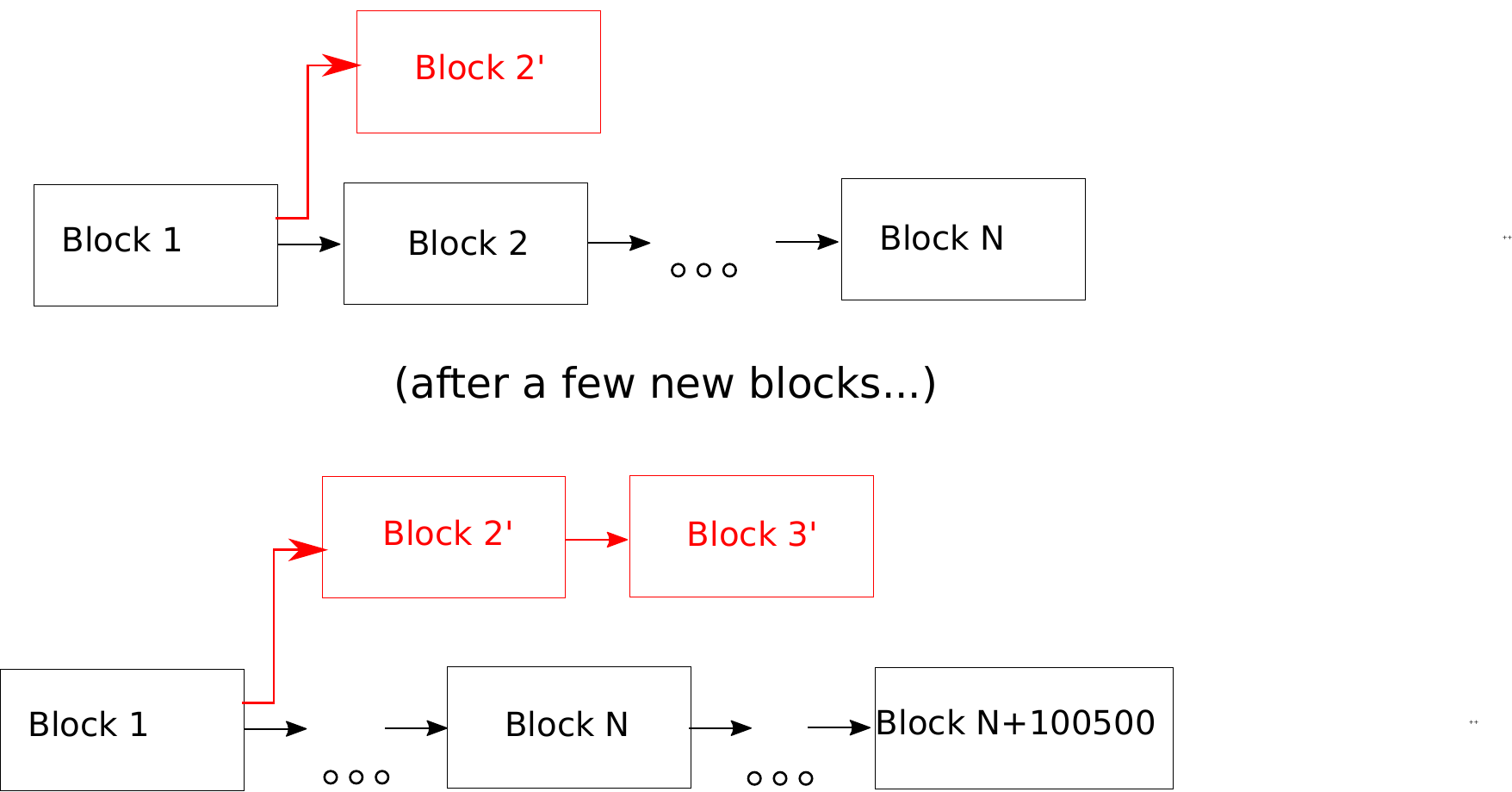}
\eeq
This is reminiscent of ER-bridge: it is non-traversable because its length grows too
fast for an observer to cross it. 
Of course, if more than a half of the users decide on a different history, it can become the 
actual history, as $(50+\epsilon)$\% of the observers(or a single observer who has 
$(50+\epsilon)\%$ of all the computational power) are in principle capable of producing a 
brand new branch and make it long. However, it will not be easy for them as they will have 
to remake all old blocks and then catch up with the main branch, which is constantly being extended by the 
rest $(50-\epsilon)\%$.

\section{Discussion}
In the previous Section we reviewed the proof-of-work-based blockchain protocol
and argued that it has many similarities to the recent proposals about complexity in quantum gravity.
We proposed “Complexity=Time” relation and used (classical) blockchain to illustrate how it already works in 
practice.

It would be interesting to make these observations more quantitative. For example, 
in blockchain a single observer who has more than $50\%$ of the total computational power 
can actually
“time travel”. An analogue to this would be the possibility of (potentially one-sided) observer 
crossing the ER bridge provided that they have enough computational power. It would be interesting 
to also relate this to Hayden--Preskill protocol.

One important distinction between modern cryptocurrency protocols and gravity complexity proposals is
that the former one is purely classical whereas the later is quantum. Quantum complexity, unlike the
classical one, is bounded and exhibits  Poincare recurrency. It would be interesting to see if
there  is a room for a switchback effect \cite{shock} in classical blockchain.
A more drastic distinction concerns observability: blockchain exploits the fact that complexity can
be easily observed: just count the number of blocks. Whereas it has been argued \cite{bouland} that quantum
complexity is very hard to determine and relating it to a rather simple observable in the bulk(volume)
have serious consequences for computer science and AdS/CFT dictionary in general.

Obviously it would be extremely interesting to quantize blockchain. There is extensive literature on "quantum blockchain",
but most of it discusses how to make classical blockchain resilient to quantum attacks.
We refer to \cite{rajan} for a
description of a "truly quantum" blockchain. 
In the context of black holes, it has been proposed that a series of shockwaves can be used 
to store information in a secure way \cite{bouland}.

Finally, it is worth noting that there are many other alternatives to proof-of-work, such as
proof-of-space\cite{proof_of_space} and proof-of-stake \cite{king2012ppcoin,saleh}. All of them try to exploit the same
idea: there is a distributed commodity(such as computational power or hard--drive space)
which is used to secure the blockchain. As long as no one has more than a half of the commodity, the 
protocol is secure.
The main disadvantage of proof-of-work is that all observers compete \footnote{In the 
actual cryptocurrency protocols this is encouraged by giving money for each created block.}
 with each other to produce(mine) blocks. This results in a huge energy consumption. Other protocols
try to alleviate this by sometimes allowing certain users to create blocks “for free”(in terms of 
resources required). The main problem with this is the so-called “nothing-at-stake” attack:
the selected observers will then try to add blocks to different branches as much as possible,
making sure they will get a reward. One has to add extra rules to penalize such behavior.

\section*{Acknowledgment}
I would like to thank C.~King for collaboration at the early stages of this project.
The author is indebted to A.~Kotelskiy for patiently explaining different aspects of cryptocurrencies
and blockchains. Also I would like to thank F.~Popov, B.~\v{S}oda and Y.~Zhao for numerous discussions and comments.
This material is based upon work supported by the Air Force Office of Scientific Research under 
award number FA9550-19-1-0360. It was also supported in part by funds from the University of California. 

I gratefully acknowledge support from the Simons Center for Geometry and Physics, Stony Brook University at which some of the research for this paper was performed.

Also I would like to acknowledge support from Berkeley
Center for Theoretical Physics and by the Department of Energy, Office of Science, Office
of High Energy Physics under QuantISED Award DE-SC0019380 while visiting Berkeley
Center for Theoretical Physics.

\printbibliography
\end{document}